\documentclass[multphys,vecphys]{svmult}

\usepackage{makeidx}         % allows index generation
\usepackage{graphicx}        % standard LaTeX graphics tool
\usepackage{multicol}        % used for the two-column index
\usepackage[bottom]{footmisc}% places footnotes at page bottom
\makeindex             % used for the subject index
\begin{document}

\title*{Multi-scale simulations of merging galaxies with supermassive black 
holes}
% Use \titlerunning{Short Title} for an abbreviated version of
 % your contribution title if the original one is too long
\author{Lucio Mayer\inst{1}, Stelios Kazantzidis\inst{2}, Piero Madau
\inst{3}, Monica Colpi\inst{4}, Thomas Quinn\inst{5} \and James
Wadsley\inst{6}}
% Use \authorrunning{Short Title} for an abbreviated version of
% your contribution title if the original one is too long
\institute{Institute f\"ur Astronomie, ETH Z\"urich at H\"onggerberg, 8093 
Z\"urich, Switzerland
\texttt{lucio@phys.ethz.ch}
\and Kavli Institute for Cosmological Physics, University of Chicago, Chicago, IL 60637, USA \texttt{stelios@cfcp.uchicago.edu}
\and Department of Astronomy and Astrophysics, University of California, 1156 High Street, Santa Cruz, CA 95064, USA \texttt{pmadau@ucolick.org}
\and Dipartimento di Fisica, Universita' Milano-Bicocca, Piazza della Scienza 3, 20126 Milano, Italy \texttt{Monica.Colpi@mib.infn.it}
\and Astronomy Department, University of Washington, Stevens Way, Seattle, WA 98195, USA \texttt{trq@astro.washington.edu}
\and Department of Physics and Astronomy, McMaster University, 1280 Main Street West, Hamilton, ON L8S 4M1 Canada \texttt{wadsley@mcmaster.ca}}
%
% Use the package "url.sty" to avoid
% problems with special characters
% used in your e-mail or web address
%
\authorrunning{Mayer et al.}

\maketitle{}

%Multi-scale simulations of merging galaxies with supermassive black 
%holes

\begin{abstract}

We present the results of the first multi-scale N-Body+SPH 
simulations of merging galaxies containing central supermassive 
black holes (SMBHs) and having a spatial resolution of only a few parsecs.
Strong gas inflows associated  with equal-mass 
mergers produce non-axisymmetric 
nuclear disks with masses of order $10^9 M_{\odot}$, resolved
by about $10^6$ SPH particles.     
Such disks have sizes of several hundred parsecs but most of their 
mass is concentrated within less than $50$ pc. 
We find that a close SMBH pair forms after the 
merger. The separation of the two SMBHs then shrinks further owing 
to dynamical friction against the predominantly gaseous 
background. The orbits of the SMBHs decay down to the
minimum resolvable scale in a few million years 
for an ambient gas temperature and density typical
of a region undergoing a starburst. These results suggest the initial
conditions necessary for the eventual coalescence of the two holes 
arise naturally from the merging of two equal-mass galaxies 
whose structure and orbits are consistent with the predictions of the 
$\Lambda$CDM model. 
Our findings have
important implications for planned gravitational wave detection 
experiments such as {\it LISA}.

\end{abstract}

\section{Introduction}
\label{sec:1}

Recent observations  of molecular gas in the nuclear region of candidate
merger remnants such as starbursting ultraluminous infrared galaxies
(ULRIGs) reveal the presence of rotating gaseous disks which
contain in excess of $10^9 M_{\odot}$ of gas within a few hundred
parsecs ([3],[5]). 
Some of these galaxies, such as Mrk 231, also host a powerful AGN.
The high central concentration of gas is likely the 
result of gaseous  inflows driven by the prodigious tidal torques and 
hydrodynamical shocks occurring during the merger ([1], [7]), and possibly 
provides the reservoir that 
fuels the SMBHs. If both the progenitor galaxies 
host a SMBH the two holes may sink and eventually coalesce as a 
result of dynamical friction against the gaseous background.
The sinking of two holes with an initial separation of $400$ pc in 
a nuclear disk described by a Mestel model has been studied by [6]
and [4].
Here we avoid any assumption on the structure of the
nuclear gaseous disk and initial separation of the pair: rather, we follow 
the entire merging process starting from
when the cores of the two galaxies are hundreds of kiloparsecs apart up 
to the point where they merge, produce a nuclear disk, and leave a pair 
of SMBHs separated by the adopted force resolution of $10$ pc. 
Previous hydrodynamical simulations
of merging galaxies with SMBHs have not followed the evolution of the 
central region below a scale of a few 
hundred parsecs due to their limited mass and force resolution ([11], [7]). In this study we are able for the
to extend the dynamic range of previous works by orders of magnitude 
using the technique of particle splitting.

\section{The Numerical Simulations}
\label{sec:2}

Our starting point are the high-resolution simulations presented in 
[7] which, as the new simulations presented here, were performed with 
the parallel tree+SPH code GASOLINE [13]. 
In particular
we refer to those simulations that followed the merger between
two equal-mass, Milky-Way sized early type spirals having 10\% of the
mass of their exponential disk in a gaseous component and the rest
in stars. The structural parameters of the two disks and their NFW dark matter
halos, as well as their initial orbits, are
motivated by the results of cosmological simulations. We apply the 
static splitting of SPH particles [8] to increase the gas mass 
resolution in such calculations (the stars and dark matter resolution
remain the same), and reduce the gravitational softening 
accordingly as we increase the mass resolution. We select an output
about $50$ Myr before the merger is completed, 
when the cores of the two galaxies are still $\sim 6$ kpc 
away, and split each SPH particle into $8$ children, reaching a mass 
resolution of about 3000 $M_{\odot}$ in the gas component (Figure 1).
The gravitational softening of the gas particles is 
decreased from $200$ pc to either $40$ pc, $10$ pc, or $2$ pc (one run is
performed for each of the three different softenings).
With the new mass 
resolution even for the smallest among the gravitational softenings 
considered the number of SPH 
particles within a sphere of radius equal to the local Jeans length is 
much larger than twice the number of SPH neighbors (=32), 
thus avoiding spurious fragmentation ([2]). 
The two SMBHs are point masses with a softening length set 
equal to $10$ pc and a mass of $3 \times 10^6 M_{\odot}$. 
The simulations presented
here do not include star formation but gas masses are only
a factor of $3$ higher than those measured in the star formation simulations
of [7] at the same evolutionary stage.

\begin{figure}
\centering
\includegraphics[height=12cm]{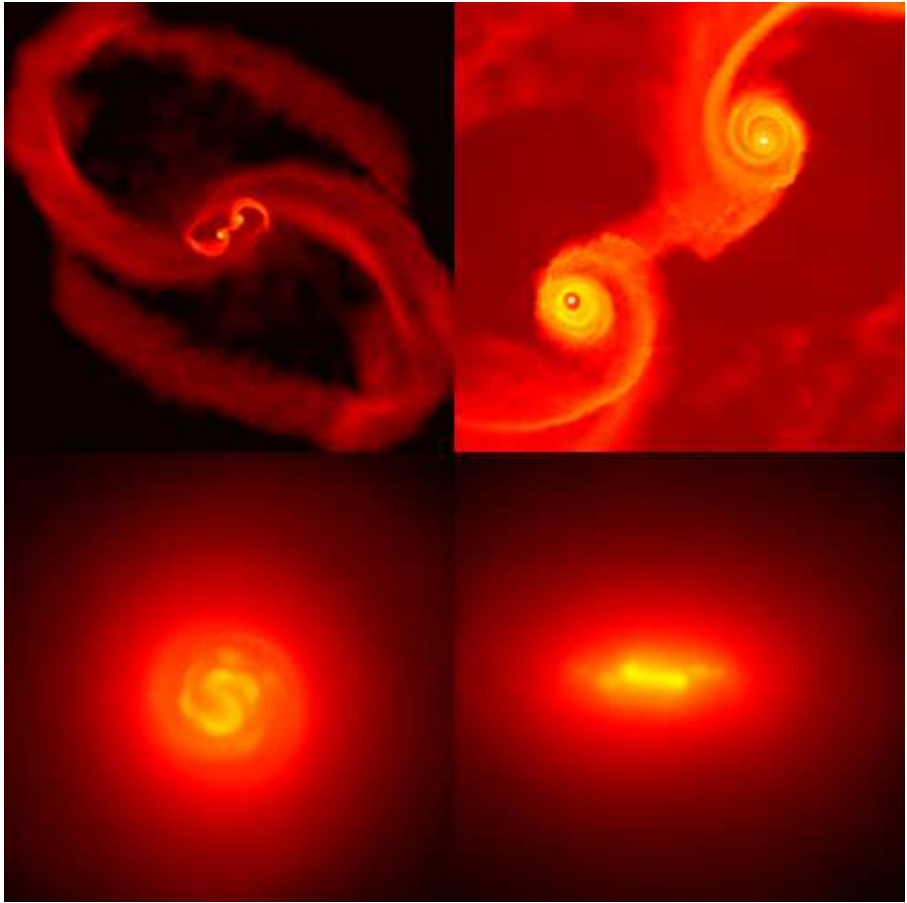}
\caption{Color coded density map of the nuclear region $50$ Myr before the merger (top panels) and just after the merger (bottom panels). The top panels show
a box $30$ kpc on a side (left) and a zoom-in within the inner $6$ kpc (right).The bottom panels show the inner 300 pc, with the disk seen face-on (left)
and edge-on (right). An adiabatic equation of state with $\gamma=7/5$ 
was used in this run.}
\label{fig:1}      
\end{figure}

We have run a suite of simulations with different 
prescriptions for the gas thermodynamics, and show here the
results of two runs in which radiative cooling and heating processes
are not included directly, rather an adiabatic equation of state
with either $\gamma=7/5$ or $\gamma=5/3$ is adopted (irreversible
shock heating, which is important during the merging phase, is
included via an artificial viscosity term in the energy equation).
According to the radiative transfer calculations of [10]
the case $\gamma=7/5$ approximates quite well the
balance between radiative heating and cooling in a starburst
galaxy (in [7] a central starburst indeed does occur 
in the final phase of the merger that we are considering here).
A stiffer equation of state such as that with $\gamma=5/3$ might instead be 
relevant when an additional strong heating source, for example AGN feedback,
comes into play ([11]).
Although one should follow directly the various cooling and heating 
mechanisms, 
this simple scheme can provide us with a guide
of how gas thermodynamics can affect the results.

\section{Results}
\label{sec:3}

Our simulations allow to assess in a self-consistent way the evolution of the 
nuclear region of a gas-rich remnant of a major merger as well as the orbital evolution
of SMBHs in the nuclei of the two merging galaxies.

\subsection{Gas Inflows and the Structure of the Nuclear Disks}

About 80\% of the gas originally belonging to the two galaxies is funneled
to the central kiloparsec during the last stage of the merger and settles into
two rotationally supported disks. When the cores of the two galaxies 
finally merge, the two disks also merge into a single gaseous core which
rapidly becomes rotationally supported as radial motions are largely dissipated
in shocks. 
The disk however remains non-axisymmetric, with evident bar-like
and spiral patterns (Figure 1). A coherent thick disk forms independent
of the relative inclination of the initial galactic disks, albeit the
orientation of its angular momentum vector relative to the global angular
momentum will change depending on those initial parameters ([7]). 
In the run with $\gamma=7/5$, that was designed  
to reproduce the
thermodynamics of a starburst region, the disk has a vertical
extent of about $20$ pc and $v_{rot}/\sigma > 1$ out to about $600$ pc.
The thickness is about $5$ times higher in the run with $\gamma=5/3$.
The scale height in the $\gamma=7/5$ run is comparable to that of 
the disks in the multi-phase simulations of a $1$ kpc-sized nuclear 
region performed by [12]. 
The simulations of [12] include radiative cooling and resolve the turbulence 
generated from  supernovae explosions as well as gravitational 
instability, suggesting that our equation of state yields a
characteristic pressure scale that accounts for the combined thermal
and turbulent pressures.

\begin{figure}
\centering
\includegraphics[height=9cm]{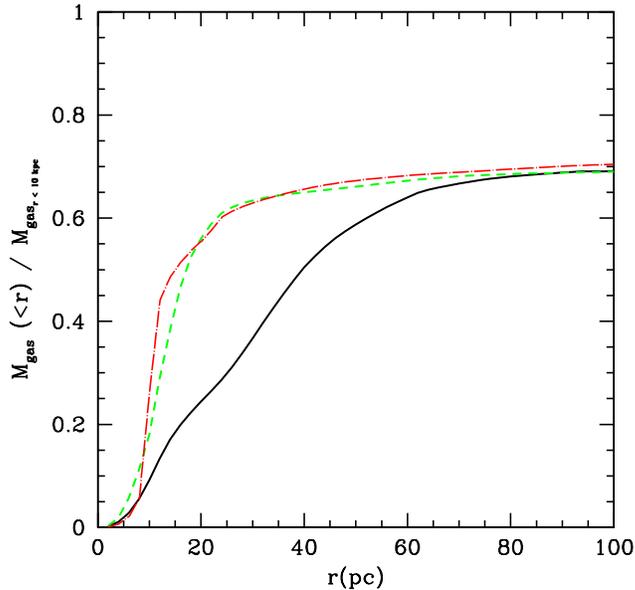}
\caption{Dependence of gas mass inflow on force resolution. The curves show
the cumulative radial gas mass profile (normalized to the total gas mass
contained within a sphere of $10$ kpc in radius) for three runs using an adiabatic equation of state with $\gamma=7/5$ and
different values of the gravitational softening of the gas particles (solid line=$40$ pc, dashed line=$10$ pc, dot-dashed line=$2$ pc).}
\label{fig:2}       
\end{figure}

Gravitational torques and the balance between gravity and the thermodynamical
pressure at small scales
depend ultimately on the adopted gravitational softening.
We verified that the mass inflow seen in the simulations for a given
value of $\gamma$ converges as the softening approaches $10$ pc (see
Figure 2). Convergence in the disk vertical extent is also observed at such
a spatial resolution.
Figure 2 shows that at high resolution more than 60\% of the mass piles up within as little as $30$ pc.

\subsection{Pairing of SMBHs}

The merger between the two galactic cores delivers a close but 
unbound pair of SMBHs. The pair is separated by
about $100$ pc and  is embedded within the
newly formed nuclear gaseous disk. Up to this point
the orbital decay of the two SMBHs had been equivalent to that of the cores in which
they were embedded, and was driven by dynamical friction of  the cores 
within the surrounding {\it collisionless} background of stars and dark matter.
Once inside the massive gaseous disk the orbital decay of the two 
holes is dominated by dynamical friction in a {\it gaseous} background 
([9]).
The intensity of the drag is then higher or lower depending on 
whether the black holes move supersonically or subsonically with respect 
to the background, and increases also as the the characteristic density of 
the background increases. The run
with $\gamma=7/5$ falls in the supersonic regime (the orbital velocity of the
black holes is of order $300$ km/s, which corresponds to a temperature of about
$10^6$ K), while the black holes move slightly subsonically in the run 
with $\gamma=5/3$. In the latter run the average background density is 
also a factor $\sim 2$ lower compared to the run with $\gamma=7/5$. These differences
in the thermal and density structure of the remnants explain why 
in the run with
$\gamma=7/5$ the two black holes reach a separation comparable to 
the force resolution limit
($10$ pc) in about $10$ Myr whereas they remain separated by a distance 
larger than $100$ pc in the other run (Figure 3).
In neither case the SMBHs form a binary
by the end of the simulation. However, with an even higher force resolution 
the  formation of a bound pair is likely in the $\gamma=7/5$ since the 
orbital energy of the binary is only marginally positive 
at $t=5.128$ Gyr. Instead in the simulation with $\gamma=5/3$ the binding
of the two SMBHs will be aborted because their orbital
decay time is longer than the Hubble time on the last few orbits.

\begin{figure}
\centering
\includegraphics[height=9cm]{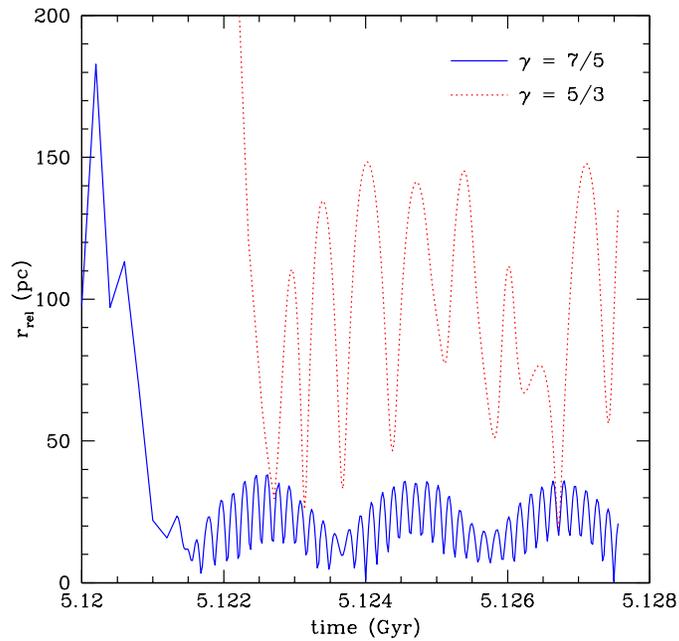}
\caption{Orbital separation of the two black holes as a function of time. Two
runs with different values of $\gamma$ and a gas gravitational softening 
of $10$ pc are shown (see labels). The curves 
start from the time at which the galaxy cores have already merged.}
\label{fig:3}      
\end{figure}

As shown in Figure 3 the two black holes end up on moderately eccentric orbits 
($e=0.3-0.5$) in all the simulations. However the orbits might 
circularize as the evolution proceeds further
([4], and these proceedings).

\section{Conclusions}
\label{sec:4}

We have performed multi-scale hydrodynamical simulations of merging galaxies
with SMBHs and
shown that dense, rotationally supported nuclear disks are the natural
outcome of dissipative mergers starting from cosmologically motivated initial
conditions.  The nuclear disks reported here likely provide the reservoir of gas that fuels the central SMBHs.
The orbital evolution of the close pair of SMBHs formed at the center 
of the merger remnant is dominated by dynamical friction against the
surrounding gaseous medium. The details of this process are extremely sensitive to thermodynamics
of the gas.
Our results indicate that the formation of a bound SMBH pair requires
an equation of state not stiffer than that expected during a major starburst. 
This suggests that either 
AGN feedback has a minor impact on the gas in the disk or that strong 
AGN feedback has to be delayed for several 
million years after the galaxy merger is completed. 
In the second case the coalescence
of the two black holes will occur when the merger remnant is a powerful
starburst, such as a ULRIG, rather than a powerful AGN.

\paragraph{Acknowledgemnts} 
SK is supported by the Swiss National Science Foundation and by The Kavli Institute for Cosmological Physics (KICP) at The University of Chicago.
We thank David Merritt for helpful comments. The simulations were performed on 
Lemieux at the Pittsburgh Supercomputing 
Center and on the Zbox and Zbox2 supercomputers at the University of Z\"urich.
%

%%%%%%%%%%%%%%%%%%%%%%%% referenc.tex %%%%%%%%%%%%%%%%%%%%%%%%%%%%%%
% sample references
% "physics"
%
% Use this file as a template for your own input.
%
%%%%%%%%%%%%%%%%%%%%%%%% Springer-Verlag %%%%%%%%%%%%%%%%%%%%%%%%%%

%
% BibTeX users please use
% \bibliographystyle{}
% \bibliography{}
%
% Non-BibTeX users please use

%%%%%%%%%%%%%%%%%%%%%%%%%%%%%%%%%%%%%%%%%%%%%%%%%%%%%%%%%%%%%%%%%%%%%%  }

%%%%%%%%%%%%%%%%%%%%%%%%%%%%%%%%%%%%%%%%%%%%%%%%%%%%%%%%%%%%%%%%%%%%%%

\printindex
\end{document}